\documentclass[12pt]{article}
\usepackage[cm]{fullpage}
\usepackage{amsmath}
\usepackage{amsfonts}
\usepackage{amssymb}
\usepackage{amsthm}
\usepackage{graphicx}
\usepackage{fancyhdr}
\usepackage{xcolor}

\definecolor{bb}{rgb}{0.3, 0.5, 1}
\definecolor{bg}{rgb}{0.1, 0.1, 0.5}
\usepackage[%
    pdftitle={The title of your letter},
    pdfauthor={Your name},
    pdfsubject={The subject of the letter},
    pdfkeywords={Any keywords},
     colorlinks=true,  
     linkcolor=bg, 
     citecolor=bg, 
     urlcolor=bg, 
     pdfnewwindow=true,
     pagebackref=true,
     filecolor=bb,
     pdffitwindow=false,     
     pdfstartview={FitH}    
]{hyperref}

\def\ba{\begin{eqnarray}}
\def\ea{\end{eqnarray}}
\def\be{\begin{equation}}
\def\ee{\end{equation}}

\def\nn{\nonumber}

\def\d{\mathrm{d}}

\def\mn{_{\mu \nu}}

\def\({\left(}
\def\){\right)}
\def\ie{{\it i.e. }}
\def\nn{\nonumber}
\def\p{\partial}
\def\dpi{(\partial \pi)^2}

\begin{document}

\vskip 0.9cm

\centerline{\Large \bf de Sitter Galileon}
\vskip 0.7cm
\centerline{\large Clare Burrage\footnote{Clare.Burrage@unige.ch},
Claudia de Rham\footnote{Claudia.deRham@unige.ch} and Lavinia Heisenberg\footnote{Lavinia.Heisenberg@unige.ch}}
\vskip 0.3cm

\centerline{\em D\'epartment de Physique  Th\'eorique, Universit\'e
de  Gen\`eve,}
\centerline{\em 24 Quai E. Ansermet, CH-1211,  Gen\`eve, Switzerland}

\vskip 1.9cm

\begin{abstract}
We generalize the Galileon symmetry and its relativistic extension
 to a de Sitter background. This is made possible by studying a
 probe-brane in a flat five-dimensional bulk using a de Sitter slicing.
The generalized Lovelock invariants induced on the probe brane
 enjoy the induced Poincar\'e symmetry inherited from the bulk,
 while living on a de Sitter geometry. The non-relativistic limit
 of these invariants naturally maintain a generalized Galileon symmetry
 around de Sitter while being free of ghost-like pathologies. We comment
 briefly on the cosmology of these models and the extension to the AdS
 symmetry as well as generic FRW backgrounds.
\end{abstract}

\vspace{1cm}

\section{Introduction}

Over 44 orders of magnitudes, from millimeter length scales (TeV
energy scales) all the way up to Cosmological scales ($\sim
10^{-32-33}$eV), the theory of General Relativity is easily the
force which has successfully passed the largest variety of tests.
 In view of these successes, modifying the theory of General
 Relativity may seem a purely academic endeavor. However, the
  unification
of gravity with the rest of the standard model will undoubtedly
 result in a modification of the theory in the UV. Furthermore,
 at the other end of the energy spectrum, the Cosmological
 Constant and the late-time acceleration of the Universe might
 be seen as a sign of the breakdown of gravity on today's
 cosmological scales.

Following this philosophy, there has been a return of interest
in theories that modify gravity in the IR. Theories of massive
 gravity in particular, could play a crucial role within the
  framework of degravitation, \cite{degravitation,
  ArkaniHamed:2002fu,Dvali:2007kt}, where the weakening of
  gravity on large distances could address the cosmological
  constant problem.

Among these models, the Dvali-Gabadadze-Porrati (DGP) scenario
has played a crucial role, as the first theory of gravity
mediated by an effectively softly massive spin-2 field (resonance),
 without leading to any ghost like pathologies (when working
 around the stable branch), \cite{DGP}. While the weakening
 of gravity in DGP is not sufficient to degravitate the vacuum
 energy, small departures from DGP are expected to provide a
 useful framework for degravitation, \cite{Dvali:2007kt}, in
 particular when extending DGP to higher dimensions,
  \cite{gigashif,cascading}. Furthermore, theories of hard mass
   gravity with no ghosts in the decoupling limit, can
   potentially degravitate small amounts of vacuum energy
    \cite{deRham:2010tw}.

Another (opposite) strategy in understanding the late-time
acceleration of the Universe relies on the idea of
self-acceleration, whereby the acceleration of the Universe
is driven by the graviton's own degrees of freedom (namely
its helicity-0 mode) rather than a cosmological constant.
While the DGP scenario allows for self-accelerating solutions,
\cite{DDG}, that branch of solutions has been proven to be
 pathological, \cite{Koyama:2005tx}. However, once again,
  small modifications of the model can also provide stable
   self-accelerating solutions, \cite{deRham:2006pe}.
 This has been explored more recently in the generalization
 of the decoupling limit of DGP, proposed by Nicolis
 {\it et. al.}, \cite{Nicolis:2008in}. The decoupling limit
 of DGP is obtained by taking the Planck scale to infinity,
 while sending the graviton mass to zero in such a way that
 the helicity-0 and -2 modes of the graviton decouple,
 \cite{nima,Nicolis:2004qq}. In that limit, the resulting
 effective theory for the helicity-0 mode $\pi$ satisfies
 two  fundamental properties: First, its action is invariant
 under the Galilean symmetry $\pi \to \pi + c+v_\mu x^\mu$.
  Second, although the decoupling limit of DGP involves
  higher derivative terms in $\pi$ it does so in such a
  way that the equations of motion remain second order
  in derivative, hence avoiding the standard Ostrogradski
  instability.
 The idea behind the ``Galileon" model proposed by Nicolis
 {\it et. al.}, is to generalize the decoupling limit of DGP
  to all the possible interactions satisfying the two
  previous properties. In four dimensions, there are five
  possible interactions that satisfy the symmetry, while
  maintaining a well posed Cauchy problem. As shown in
  \cite{Nicolis:2008in,Silva:2009km}, for a certain class
   of parameters, the model allows stable self-accelerating
   solutions.
 Interestingly, the Galileon family of interactions, has
  been shown to arise generically in other theories where
  gravity is mediated by a massive spin-2 field (hard mass
  graviton) which show no signs of the Boulware-Deser ghost
   in the decoupling limit
   \cite{deRham:2010ik,deRham:2010kj,deRham:2010gu,aux1}.

Whether the Galileon can tackle the cosmological constant
problem or provide a viable phenomenology for
self-acceleration, is still an open question, but it does
provide a unique framework for the study of a scalar field
that could play a role on cosmological scale while
exhibiting the Vainshtein mechanism on small distance
scales near sources, \cite{Vainshtein:1972sx}.  As such,
 there has been a large amount of interests in the cosmology
 and the phenomenology of such models, \cite{Chow:2009fm,generic,Hinterbichler:2009kq,Creminelli:2010ba,Burrage:2010cu,Deffayet:2010qz}. However, in its original derivation from a decoupling
 limit where interactions with gravity are suppressed,
 the ``Galileon" should really be though as a scalar
 field living around flat space-time. To bring the Galileon
 a step forward and understand its behaviour around generic
 geometries, in particular around cosmological backgrounds,
 it is essential to understand its potential covariantization.

When promoting the Galileon to curved backgrounds, Deffayet
{\it et.al.} have ensured that the absence of ghosts was
preserved, \cite{Deffayet:2009wt}. However this comes at
the price of breaking the Galileon symmetry, which is only
valid on a flat geometry. The ``Covariant Galileon" can
also be derived in a natural way by embedding a probe brane
 in a five-dimensional space-time and inducing the Lovelock
  invariants on the brane, \cite{deRham:2010eu}. The Galileon
  symmetry around Minkowski is then a simple consequence of
  five-dimensional Poincar\'e in the bulk, but it is clear
  that for an arbitrary bulk geometry the amount of symmetry
   is reduced. The symmetry enjoyed by the Galileon on an
   arbitrary background, should then in principle be
   expressed after deriving the exact bulk geometry.

In this work, we follow the same philosophy as in
\cite{deRham:2010eu} and consider the specific case where
the bulk geometry remains Ricci flat, while the brane geometry
is (anti-) de Sitter. This is made possible by working in
 the de Sitter slicing of five-dimensional Minkowski, where
 the geometry on a brane at fixed position along the extra
 dimension is (anti-) de Sitter. Looking at the five-dimensional
  Poincar\'e symmetry in the de Sitter slicing, we can
  easily generalize the ``flat" Galileon symmetry to a
  ``de Sitter" equivalent. Not only is this framework most
   useful for cosmology, and in particular for inflation
   and late-time acceleration, but equally importantly,
   this framework can be used as a playground where
   corrections from the geometry to the Galileon symmetry
   and interactions can be understood. As such the de
   Sitter Galileon provides an important generalization
   of the Galileon symmetry around a curved background.

The rest of the paper is organized as follows: We begin in
section \ref{sec:symmetries} by presenting the full Poincar\'e
 symmetry when working in the de Sitter slicing of Minkowski.
 We also discuss the implications for the non-renormalization
 theorem which is essential for the viability of the effective
  field theory of this model. We then move onto the
  probe-brane picture in section \ref{sec:Framework} and
   describe the Lovelock invariants induced on the brane.
   By taking the ``non-relativistic" limit, which corresponds
   here to a weak field limit, we recover the generalization
   of the Galileon, now valid around de Sitter. We also
   present the anti-de Sitter (AdS) equivalent of the
   Galileon and discuss its possible generalization on FRW.
We then comment on the consequences for cosmology and
specifically for inflation and late-time acceleration,
before exploring further directions in section \ref{sec:conclusion}.

\section{Symmetries and Non-Renormalization theorem}
\label{sec:symmetries}
\subsection{de Sitter Slicing of Minkowski}

Following the same philosophy as in \cite{deRham:2010eu},
 we study the brane position modulus
of a relativistic probe brane embedded in a five-dimensional
bulk. We start with describing the five-dimensional
Minkowski bulk in a de Sitter slicing and expressing the
full Poincar\'e symmetry in this gauge,
\ba
\label{eq:deSitterslicing}
\d s^2_5=e^{-2 H y}\(\d y^2+q\mn \d x^\mu \d x^\nu\)\,,
\ea
where $q\mn$ is the de Sitter metric,
\ba
\d s_{dS}^2=q\mn \d x^\mu \d x^\nu=-n^2(t)\d t^2+a^2(t)\d x^2
\ea
with $\frac{\dot a}{n a} ={\rm const}= H$.

The equivalent of the translation (parameter $c$),
rotations ($v_i$) and boost ($v_0$) involving the $y$
direction are given by
$x^a \to x^a +\delta x^a$ with
\ba
\label{dy}
\delta y &=& e^{H y}\(a\(c+v_i x^i +\frac{v_0}{2 H}+\frac 12 v_0 H \delta_{ij}x^i x^j\)-\frac{v_0}{2 a H}\)\\
\delta t&=& \frac{e^{H y}}{n}\(a\(c+v_i x^i +\frac{v_0}{2 H}+\frac 12 v_0 H \delta_{ij}x^i x^j\)+\frac{v_0}{2 a H}\)-\frac{1}{n}\(c+\frac{v_0}{H}+v_i x^i\)\hspace{20pt}\\
\delta x^i& = &\(c\, H \, x^i+v_0 x^i+\frac{v^i}{2 H}+v_j x^j x^i-\frac 12 v^i x^j x_j-\frac{v^i}{2 H a^2}-\frac{e^{H y}}{a}\(\frac{v^i}{H}+v_0 x^i\) \)\,.\hspace{20pt}
\label{dxi}
\ea
We could have expressed the equivalent of translations
along $y$ in the simpler form $\delta y=n\,  \delta t=c\,
 e^{H y} a(t)$ and $\delta x^i=0$, but have chosen the
 specific form of the transformation which reproduces
 the standard translation and rotations in flat space
 in the limit when $H\to 0$ with the lapse $n=1$
\ba
y&\longrightarrow & y+c+v_\mu x^\mu\\
x^\mu&\longrightarrow & x^\mu-v^\mu y\,.
\ea

For a brane localized at $y=\pi(x^\mu)$, the 5d dimensional
 Poincar\'e symmetry implies the following symmetry under
 shifting and boosting of $\pi$:
\ba
\label{transformation}
\pi \to \pi +\( \delta y-\delta x^\mu \partial_\mu \pi\)\Big|_{y=\pi(x^\mu)}\,.
\ea
Provided that all the brane quantities transform as scalar,
 the induced action should then be invariant under the
 transformation \eqref{transformation}.

\subsection{Non-renormalization theorem}

Let us first review the non-renormalization theorem as
presented in Ref.~\cite{Nicolis:2004qq}. Starting with
the now standard form of the Galileon,
\ba
\mathcal{L}=-(\p \pi)^2+\frac{1}{\Lambda^3}(\p \pi)^2 \Box \pi\,,
\ea
and expanding around a classical solution,
$\pi=\pi_{\rm cl}+\varphi$, the leading order Lagrangian
for the perturbations around that background is
\ba
\mathcal{L}_{\varphi}=Z\mn(\pi_{\rm cl})\p^\mu \varphi \p^\nu \varphi\,,
\ea
with
\ba
Z\mn=-(1+2\frac{\Box \pi_{\rm cl}}{\Lambda^3})\eta\mn-\frac{2}{\Lambda^3}\p_\mu \p_\nu \pi_{\rm cl}\,.
\ea
To simplify the argument, let us assume a simpler kinetic
term of the form $Z\mn=-Z \eta\mn$,
\cite{Nicolis:2004qq}\footnote{Sufficiently close to a
source all the non-zero elements of $Z_{\mu\nu}$ will
indeed become of the same order} , with
$Z=1+2\Box \pi_{\rm cl}/\Lambda^3$, so that the canonically
normalized field $\hat \varphi$ can easily be expressed in
terms of $\varphi$, $\hat \varphi =\sqrt{Z}\varphi$, and the
Lagrangian around a non-trivial configuration is
\ba
\mathcal{L}_{\hat \varphi}=-(\p \hat \varphi)^2+\frac12 \left[\frac{(\p Z)^2}{2Z^2}-\frac{\p^2 Z}{Z}\right]\hat \varphi^2\,,
\ea
giving rise to an effective mass term for $\varphi$.
The Coleman-Weinberg 1-loop effective action is then of the form
\ba
\mathcal{L}_{\rm 1-loop}&=&(V''(\hat \varphi))^2\log\frac{V''}{M_\star^2} \nn\\
\label{CT}
&\supset&\left\{ \frac{(\p Z)^4}{Z^4} , \frac{(\p Z)^2}{Z^2}\frac{\p^2 Z}{Z}, \frac{(\p^2 Z)^2}{Z^2}\right\}\log\frac{V''}{M_\star^2}\,,
\ea
where $M_\star$ is the cutoff scale, with $M_\star \gg \Lambda$.
Within the strong coupling region the higher interactions dominate,
 $\p^2 \pi_{\rm cl}\gg \Lambda^3$, and one might worry that
  since $Z$ is large, the effective field theory goes out of
  control. However as is explicit in \eqref{CT}, the counter
  terms do not depend on the absolute value of $Z$ but only
  on the ratio with at least one of its derivatives. Quantum
  corrections can thus be under control as long as $\p Z \ll Z$
  even though the classical solution has $Z\gg 1$. In other
  words, one can have large operators,
  $\p^2 \pi_{\rm cl}\gg \Lambda^3$ as long as the effective
  theory can be reorganized so as to define new irrelevant
  operators that carry extra gradients,
  $\p^3 \pi_{\rm cl}\ll \Lambda^4$, \ie $\p/\Lambda\ll 1$.
Furthermore a key ingredient of the non-renormalization
theorem is the nature of the new counter terms. It is
clear from the form of the one-loop effective action, that
any counter term, will arise with extra gradients as compared
to the original interactions, $\p Z \sim \p^3 \pi$. This
implies that the Galileon interactions themselves will not
be renormalized, and the one-loop effective action will
only affect higher derivative terms.

Having summarized the argument in the usual Galileon scenario,
 it is straightforward to apply the same philosophy when
 working around de Sitter. In this case, the Galileon
 interaction, as we shall see later, is dressed with corrections
 coming from the Hubble parameter, schematically
 $\p \pi \to \p \pi+\beta H \pi$, with a given constant
 $\beta$ so that for instance,
  $(\p \pi)^2 \Box \pi \to (\p \pi)^2 \Box \pi +6H^2 \pi (\p \pi)^2 -8 H^4 \pi^3$. In this framework, the Hubble parameter plays a similar role
  to the gradient, $H\sim \p$, so that the previous
  counting remains identical. As a consequence, the
  de Sitter Galileon interactions are again not
  renormalized, and the effective field theory remains
   under control as long as
    $\partial /\Lambda \sim H/\Lambda \ll 1$, \ie $\p^3 \pi_{cl}/\Lambda^4 \sim H^3 \pi_{cl}/\Lambda^4 \ll 1 $ even  around classical
    configurations with
$\p^2 \pi_{cl}/\Lambda^3 \sim H^2 \pi_{cl}/\Lambda^3 \sim 1 $.

\section{de Sitter Galileon}
\label{sec:Framework}
We use the same notation as \cite{Nicolis:2008in}, with
$\Pi\mn=D_\mu D_\nu \pi$ where the covariant derivative
is taken w.r.t. the 4D metric $q\mn$ and square brackets
 $[...]$ represent the trace (w.r.t. $q\mn$) of a tensor.
 Furthermore we also introduce the following notation
\ba
[\phi^n]\equiv\p \pi \,.\, \Pi^n \,.\, \p \pi\,,
\ea
so in particular $[\phi]=D_\mu D_\nu\pi \p^\mu \pi \p^\nu \pi=q^{\mu\alpha}q^{\nu\beta}D_\mu D_\nu \pi \p_\alpha\pi \p_\beta \pi$.
We also express the Lorentz factor as
\ba
\gamma=\frac{1}{\sqrt{1+\dpi}}
\ea
with $(\p \pi)^2=q^{\mu\nu}\p_\mu \pi \p_\nu \pi$.

\subsection{Probe Brane in de Sitter Slicing}

We consider a probe brane embedded in a flat, 5-D Minkowski
 bulk, which we take to be in a de Sitter slicing
 (\ref{eq:deSitterslicing}).  The brane is positioned at
 $y=\pi(x_{\mu})$.  From the five-dimensional theory, five
 invariant quantities can be induced on the brane
 \cite{deRham:2010eu}: The tadpole ${\mathcal{L}}_1$, the
 DBI equivalent ${\mathcal{L}}_2$, the extrinsic curvature
  ${\mathcal{L}}_3$, the induced Ricci tensor ${\mathcal{L}}_4$
  and finally the boundary term from the Gauss-Bonnet
  curvature in the bulk ${\mathcal{L}}_5$. These are
  constructed out of the induced metric at $y=\pi (x^\mu)$
\ba
g\mn=e^{-2 H \pi}\(q\mn +\p_\mu \pi \p_\nu \pi\)\,.
\ea
It is convenient to remind ourselves first of their
expression around flat space-time as derived in
\cite{deRham:2010eu},
\ba
{\mathcal{L}}_2^{f}&=&\gamma^{-1}\label{eq:L2f}\\
{\mathcal{L}}_3^{f}&=&\([\Pi]-\gamma^2 [\phi]\)\\
{\mathcal{L}}_4^{f}&=&\gamma\(  \([\Pi]^2-[\Pi^2]\)+2 \gamma^{2}\([\phi^2]-[\Pi] [\phi]\)\)\\
{\mathcal{L}}_5^{f}&=&\gamma^2 \Big(
 \([\Pi]^3+2[\Pi^3]-3[\Pi][\Pi^2]\)+6\gamma^2 ([\Pi][\phi^2]-[\phi^3])\label{eq:L5f}\\
&&\hspace{20pt}-3\gamma^2([\Pi]^2-[\Pi^2])[\phi]
\Big)\,. \nn
\ea
When working around a de Sitter background, the previous
Lagrangians do not respect the full Poincar\'e symmetry.
For a de Sitter geometry   the   invariants are
\ba
\label{L1}
\sqrt{-q}{\mathcal{L}}_1^{\rm dS}&=&\frac{1}{5H}\sqrt{-q} \(e^{-5 H \pi}-1\)\\
\sqrt{-q}{\mathcal{L}}_2^{\rm dS}&=&\sqrt{-g}=e^{-4H \pi } \sqrt{-q}{\mathcal{L}}_2^{f}\\
\sqrt{-q}{\mathcal{L}}_3^{\rm dS}&=&-\sqrt{-g}K=e^{-3H \pi } \sqrt{-q}\({\mathcal{L}}_3^{f}+4 H \gamma {\mathcal{L}}_2^{f}\)\\
\sqrt{-q}{\mathcal{L}}_4^{\rm dS}&=&\sqrt{-g}R=e^{-2H \pi } \sqrt{-q}\({\mathcal{L}}_4^{f}+6 H \gamma {\mathcal{L}}_3^{f}+12 H^2 \gamma^2 {\mathcal{L}}_2^f\)\\
\sqrt{-q}{\mathcal{L}}_5^{\rm dS}&=&-\frac 32 \sqrt{-g}\, \mathcal{K}_{GB}\nn\\
&=&e^{-H \pi } \sqrt{-q}\({\mathcal{L}}_5^{f}+6 H \gamma {\mathcal{L}}_4^{f}+18 H^2 \gamma^2 {\mathcal{L}}_3^f+24 H^3 \gamma^3 {\mathcal{L}}_2^f\)
\label{L5}
\ea
where the ${\mathcal{L}}_i^{f}$ are the flat space invariants
 (\ref{eq:L2f}-\ref{eq:L5f}) and $\mathcal{K}_{\rm GB}$ is
 the Gibbons-Hawking-York boundary term associated with a
 bulk Gauss-Bonnet term $\mathcal{R}_{\rm GB}$
 \cite{Davis:2002gn,deRham:2010eu},
\ba
\mathcal{R}_{\rm GB}&=&R^2-4 R\mn^2+R_{\alpha\beta\mu\nu}^2\\
\mathcal{K}_{\rm GB}&=&-\frac 23 K\mn^3+K K\mn^2-\frac 13 K^3-2G\mn K^{\mu\nu}\,.
\ea
We have used the following expressions for the extrinsic
 and intrinsic curvature on the brane,
\ba
K\mn&=&\gamma e^{-H\pi}\(\Pi\mn +H\gamma\mn+H\p_\mu \pi \p_\nu \pi  \)\\
R\mn&=&2 H \gamma^2 \Pi\mn+\gamma^2 ([\Pi]\Pi\mn-\Pi^2\mn)+\gamma^4(\phi^2\mn-[\phi]\Pi\mn)\\
&&+H\gamma^2 \(3 H+\gamma^2 [\Pi]+\gamma^2([\Pi](\p \pi)^2-[\phi])\)(q\mn+\p_\mu\pi\p_\nu\pi)\nn\\
R&=&e^{2 H \pi}\(12 H^2 \gamma^2 +6 H \gamma^2 \([\Pi]-\gamma^2[\phi]\)+\gamma^2([\Pi]^2-[\Pi^2])
+2 \gamma^4 ([\phi^2]-[\phi][\pi])\)\nn\,.
\ea
One can check explicitly that the Lagrangians
(\ref{L1}-\ref{L5}) defined in this way  are symmetric
under the transformation (\ref{dy}-\ref{dxi}).
Furthermore they also satisfy the recursive relations
related to the `Universal Field Equations', first introduced
 by Fairlie {\it et. al.}, \cite{Fairlie:1991qe}, and
 explained more recently within the context of the
 Galileon \cite{Fairlie:2011md,deRham:2010eu}
\ba
\label{recursive}
\mathcal{L}^{\rm dS}_{n+1}=-e^{H \pi}\gamma^{-1}\frac{\delta \mathcal{L}^{\rm dS}_n}{\delta \pi} \hspace{20pt}{\rm for}\hspace{10pt} n\ge 1\,,
\ea
which generalize the flat space-time relations.  It is
 straightforward to check that there cannot be any further
  invariant beyond $n=5$ because
  $\delta_\pi \mathcal{L}^{\rm dS}_5$ is a total derivative.

\subsection{Non-relativistic Limit}
When building the action for this Galileon field the
Lagrangian ${\mathcal{L}}_2^{\rm dS}$ will enter with a
coefficient $\lambda$ related to the tension on the brane,
 $S\supseteq \int d^4x\;\lambda {\mathcal{L}}_2^{\rm dS}$.
We can build the non-relativistic limit, $(\partial \pi)^2 \ll 1$,
 of this theory by canonically normalizing the field
  $\pi=\hat \pi/\sqrt{\lambda}$ and sending $\lambda\to \infty$.
   In this limit the Galileon symmetry becomes (after setting
   the lapse to $n=1$),
\ba
\label{dS_galileon}
\hat \pi \longrightarrow \hat \pi + e^{H t}\(c+v_i x^i+\frac 12 v_0 H x^i x_i\)+\frac{v_0}{H}\sinh (Ht)\,,
\ea
which in the flat space limit $H\to 0$ reduces to the
Galileon shift symmetry
$\hat \pi \to \hat \pi + \(c+v_\mu x^\mu\)$ of Nicolis
{\it et. al.} \cite{Nicolis:2008in}. Keeping the lapse
arbitrary to retain the gauge freedom, the Galilean
transformation becomes,
\ba
\label{dS_galileon_cov}
\hat \pi \longrightarrow \hat \pi +a(t)\(c+v_i x^i+v_0 H x^i x_i-v_0\(\frac{n(t)}{\dot a(t)}\)^2\)\,,
\ea
with $\frac{\dot a}{a n}=H=$const.

The specific combination of ${\mathcal{L}}^{\rm dS}_1$
 and ${\mathcal{L}}^{\rm dS}_2$ which remains finite in
  the limit $\lambda \to \infty$ is
\ba
{\mathcal{L}}^{({\rm NR})}_2=\lambda({\mathcal{L}}^{\rm dS}_2+4 {\mathcal{L}}^{\rm dS}_1- 1) &  \xrightarrow{\ \lambda\, \to\, \infty \ } & \frac12\, \((\p \hat\pi)^2-4 H^2 \hat \pi^2\)\,,
\ea
which is  indeed  invariant under the non-relativistic
transformation \eqref{dS_galileon}, and which gives back
the usual first Galileon  kinetic term, $1/2(\partial \pi)^2$,
in the limit $H\to 0$.

We can now go further and consider the non-relativistic limit
 of the higher order invariants (extrinsic curvature term,
 scalar curvature, etc...) which are also invariant under
 \eqref{dS_galileon}. The ``de Sitter" generalization of
 the Galileon derivative interactions are then:
\ba
\label{eq:galileonNR2}
{\mathcal{L}}^{({\rm NR})}_2&=&(\p \hat\pi)^2-4 H^2 \hat \pi^2\\
{\mathcal{L}}^{({\rm NR})}_3&=& (\p \hat\pi)^2 \Box \hat \pi+6 H^2 \hat \pi (\p \hat\pi)^2-8 H^4 \hat \pi^3\\
{\mathcal{L}}^{({\rm NR})}_4&=& (\p \hat\pi)^2\([\hat \Pi^2]-[\hat \Pi]^2\)-6 H^2 \hat \pi (\p \hat\pi)^2 \Box \hat \pi \\
&&-\frac 12 H^2 (\p \hat\pi)^4-18H^4 \hat \pi^2 (\p \hat\pi)^2+12 H^6 \hat \pi^4\nn\\
{\mathcal{L}}^{({\rm NR})}_5&=&(\p \hat\pi)^2\([\hat \Pi]^3-3[\hat \Pi^2][\hat \Pi]+2 [\hat \Pi^3]\)-4 H^2 \hat \pi  (\p \hat\pi)^2\([\hat \Pi^2]-[\hat \Pi]^2\) \label{eq:galileonNR5}
\\ && -\frac 12 H^2 (\p \hat\pi)^4 \Box \hat \pi \nn+12 H^4\hat \pi^2 (\p \hat\pi)^2 \Box \hat \pi+2 H^4 \pi (\p \hat\pi)^4
\\ && +24 H^6 \hat \pi^3  (\p \hat\pi)^2-\frac{48}{5}H^8 \hat \pi^5\,.
\nn
\ea
One can then check that the actions
$S^{({\rm NR})}_i=\int a^3(t) n(t) {\mathcal{L}}^{({\rm NR})}_i$
 are indeed invariant under the transformation
 \eqref{dS_galileon_cov} up to a total derivative. In flat
 space-time, in addition to these invariants, there is also
 a tadpole ${\mathcal{L}}^{({\rm NR})}_1=\pi$ which enjoys the
 accidental Galileon symmetry. More precisely, under a Galileon
 transformation, the tadpole is invariant up to a constant term
 which is irrelevant around flat space-time. In de Sitter,
 however this constant would gravitate, and affect the equations
 of motion, so strictly speaking around a curved background,
 there is no equivalent of the tadpole contribution. However
 it is worth pointing out that, in its fully relativistic form,
 ${\mathcal{L}}_1^{(\rm dS)}$ in \eqref{L1} does satisfy the
  relativistic symmetry \eqref{transformation} up to a total
  derivative, with no constant left over, so in the full picture,
  the tadpole remains an invariant at the same level as the other
  interactions.
\subsection{Anti-de Sitter}
\label{sec:AdS}
To recover the symmetries for the Galileon living in AdS,
 we can embed a 4d AdS brane in an extra dimension of time
  rather than space as performed previously. We emphasize
  that this is only a standard ``trick" to recover the AdS
   symmetry in 4d, and by no means do we consider the higher
   dimensional scenario physical. This is equivalent as to
    perform the following Wick-rotation, $y \to i y$ which
    implies,
\ba
\begin{array}{ccc}
{\rm dS\ variables}  &\longrightarrow &  {\rm AdS\ variables} \\
\{y, t, x^3\} &\longrightarrow & i \{y,- z,t\}\\
\{\pi, H, K, \mathcal{K}_{GB}\} &\longrightarrow & i \{\pi, -H, K, -\mathcal{K}_{GB}\}\\
\mathcal{L}_{n}^{\rm dS} &\longrightarrow & (-i)^n\mathcal{L}_{n}^{\rm AdS}
\end{array}
\ea
More specifically, we start with the AdS slicing of
five-dimensional Minkowski,
\ba
\d s^2 = e^{-2 H y}\(-\d y^2+\d z^2+e^{-2 H z}\eta_{\alpha\beta}\d x^\alpha \d x^\beta\)\,,
\ea
where in this section the indices $\alpha, \beta$ run over
time and two of the space directions, $\alpha, \beta =0, 1, 2$,
 with coordinates $x^\alpha=\{t,x^1,x^2\}$. These indices are
  raised and lowered using the $1+2$-dimensional Minkowski
  metric $\eta_{\alpha \beta}$.
In this gauge, the equivalent of the translations and boosts
 involving the extra dimensions are $x^a\to x^a+\delta x^a$
 with
\ba
\delta y&=&(c+v_\alpha x^\alpha)e^{H(y-z)}-\frac{v_z}{H}\partial_z g(y,z)\\
\delta z &=& (c+v_\alpha x^\alpha)(1-e^{H(y-z)})-v_z g(y,z)\\
\delta x^\alpha&=& cH x^\alpha+v^\alpha \(\frac{e^{H z}}{H}(e^{Hy}-\sinh Hz)-\frac{H}{2}x_\alpha x^\alpha\)\\
&&+x^\alpha H v_\beta x^\beta
-v_z x^\alpha(1-e^{H (y+z)})\nn\,,
\ea
with $g(y,z)=\frac 1 H (1- e^{H y}\cosh H z)+\frac H 2 e^{H(y-z)}x_\alpha x^\alpha$.
Once again, we then consider a brane localized at $y=\pi(x^\mu)$,
and the resulting symmetry on the brane is given by
\eqref{transformation}. The invariants on the brane,
 are then easily expressed as in (\ref{L1}-\ref{L5}).
 In particular,
\ba
{\mathcal{L}}_1^{\rm AdS}&=&\frac{1}{5H} \(e^{-5 H \pi}-1\)
\ea
and the other invariants are expressed in terms of
${\mathcal{L}}_1^{\rm AdS}$ using a similar `Universal
Field Equations' recursive relation as \eqref{recursive},
\ba
\label{recursiveAdS}
\mathcal{L}^{\rm AdS}_{n+1}=-e^{H \pi}\tilde \gamma^{-1}\frac{\delta \mathcal{L}^{\rm AdS}_n}{\delta \pi} \hspace{20pt}{\rm for}\hspace{10pt} n\ge 1\,,
\ea
with
\ba
\tilde \gamma=\frac{1}{\sqrt{1-(\p \pi)^2}}\,,
\ea
where now all contractions are performed with respect to
the background AdS metric.

\subsection{FRW}

As such, the symmetry we have inferred is only preserved
on a pure (anti)- de Sitter background. However, for any
cosmological scenario, we would need to include at slow-roll
departures from de Sitter. One could in principle covariantize
this model, similarly to that performed in \cite{Deffayet:2009wt},
 however we would then loose the Galileon symmetry. Instead,
 another way forward is to generalize the description to a
  generic Friedmann-Robertson-Walker (FRW) background.
In principle, this extension is straightforward when
 considering the FRW-slicing of five-dimensional Minkowski
\ba
\d s^2&=&e^{-2 H y}\Big[a^2(t)\d x^2+\d y^2+2(1+H y-e^{Hy})\frac{\dot H}{n H^2}\d y \d t\\
&&-\big(1-\frac{\dot H}{n H}\big)\(1+2 (1-e^{Hy})\frac{\dot H}{n H^2}+y \frac{\dot H}{n H}\)\d t^2
\Big]\nn\,,
\ea
however the realization of the Poicar\'e symmetry in this
gauge is highly non-trivial, and the its non-relativistic
limit appears non-local. This is however not surprising,
as there is no reason why FRW which is not maximally
symmetric should enjoy similar amount of symmetry. In
 the five-dimensional picture, the coordinate transformation
  to transfer from flat slicing of Minkowski to an FRW
  slicing is non-local, and so the Poincar\'e symmetry
  expressed in the five-dimensional FRW slicing does not
   have the similar close form as in de Sitter.

\section{Outlooks}
\label{sec:conclusion}

The Galileon theory, as formulated in flat space, was
particularly interesting for cosmology because the Galileon
symmetry generalises the shift symmetry  required (at least
in a softly broken form) for a successful realisation of the
 simplest models of inflation \cite{Burrage:2010cu}.
 However, by definition, inflation does not occur in flat
  space and so the flat space Galileon symmetry will be
   broken during inflation.  While inflation is occurring
   the Universe will look like a realisation of a de Sitter
   geometry, (with some corrections so as to gracefully exit
    and let the Universe reheat).  The de Sitter geometry
    is then a very good approximation whilst inflation is
    taking place, and therefore we can ask whether the
    generalisation of the Galileon symmetry to a de Sitter
    background gives a theory that is suitable for inflation.

From the presence of the tachyonic mass term in
\eqref{eq:galileonNR2} it is clear that the Galileon will
suffer from an instability on time scales proportional to
the Hubble time and that there will be problems with
attempting to build models of inflation satisfying the
 slow roll conditions  $\ddot{\pi}\ll3H|\dot{\pi}|\ll3M_PH^2$.
 It might seem possible at first to combine several of the
 Galileon interactions, so as to construct a suitable
 potential with a stable minimum for $\pi$, however the
 existence of the shift symmetry indicates that any
 configuration is subject to decay. In flat space,
 the Galileon respects a shift symmetry $\pi \to \pi +c$
 which is precisely what is required for inflation.  On de
  Sitter, however, that symmetry is promoted to
   $\pi \to \pi +a(t)c$, which corresponds to an
   unstable mode.  Therefore even if a slow roll
   solution, $\pi_{\rm inflate}(t)$, can be found
   it will be unstable to decay into $\pi_{\rm inflate}+a(t)c$.
   This cannot obey the slow roll conditions for more than
   a few e-foldings unless $c$ is tuned to be very small.
   Since all values of $c$ will be generated by the
   symmetry this is not a natural scenario.

This does not prove, however, that there are large
corrections to the inflationary models built around the
flat space Galileon theory.  What we consider here is a
disjoint scenario, although the motivations are similar.
In this article we consider a scalar field scenario for
which the full de Sitter Galileon symmetry is realized.
For the flat space Galileon inflationary model
\cite{Burrage:2010cu}, the Galileon symmetry is softly
broken allowing slow roll inflation.  A coupling to the
 non-flat background geometry will induce, via quantum
 effects, small corrections of order $H/\Lambda$ but
 these will not act to drive the theory towards one in
 which the de Sitter Galileon symmetry is realized.  A
 straightforward way to see that these two theories are
 qualitatively different is to consider the five-dimensional
 brane world parent scenarios.  The orientation of the branes
 in the two scenarios is totally different; corresponding to
 arranging the brane to lie along the fifth dimension of either
 a flat or de Sitter slicing of the bulk geometry.  Therefore
 one scenario is not a small perturbation of the other unless
 $H$ is vanishingly small, which is certainly not the case
 during inflation.

In some scalar field models, such as DBI, inflation is
still possible even though the scalar mode is not slow
rolling. In the non-relativistic (Galileon) version of
this theory this is not possible, as the unstable mode
will push the model outside the inflationary region. In
the relativistic realization, however, the possibility
might still be open.
Another alternative, is to realize inflation in a de Sitter
 Galileon model by breaking the symmetry.  For example  by
 introducing a positive mass term, which would flip the sign
 of the tachyonic mass $-4H^2 \pi^2$ in \eqref{eq:galileonNR2},
  or by considering the tadpole term
  ${\mathcal{L}}^{({\rm NR})}_1=\pi$.
Then it would be possible to  obtain a slowly rolling-scalar
field, although the size of the symmetry breaking term would
 have to be fine tuned. As shown in \cite{Burrage:2010cu},
  although such terms would break the symmetry, they would
   not lead to any counter-terms and would themselves not
   be renormalized from the other Galileon interactions.

One of the initial motivations for studying Galileon
theories was the possibility to explain the late time
accelerated expansion of the Universe through a self
accelerated solution.  A slow roll solution in the de
Sitter Galileon theory may also explain the late time
accelerated expansion  as a form of dark energy.  The
instability present in the de Sitter Galileon, although
disastrous during inflation, would actually be acceptable
 in the late-time de Sitter expansion. In that case the
 instability would only manifest itself over periods of
 time comparable to the age of the Universe. In such a
 model, we would have a small time variation of dark
 energy, due to a very light scalar field on cosmological
  scale. On solar system scales, however, the usual
  Vainshtein mechanism would take over, evading the
  usual fifth force constraints, \cite{Vainshtein:1972sx}.

The flat space Galileon interactions contain only derivatives
of $\pi$.  In contrast the de Sitter Galileon interactions (\ref{eq:galileonNR2})-(\ref{eq:galileonNR5}) contain additional,
potential-like, terms which are functions of $\pi$ and terms
 which a mixtures of $\pi$ and gradients of $\pi$. One way to
 ensure that no worse instability exist, and to have solutions
 suitable for dark energy is to work near extrema of the
 potential-like contributions.  In that case, we have to
 limit ourselves to a subset of possible choices of coefficients
  for the ${\mathcal{L}}^{({\rm NR})}_i$ which have extrema of
   the potential  within the regime of validity of the
   effective field theory, $\pi \ll \Lambda^4/H^3$, with
   the positive vacuum energy measured today.

Finally, it is worth commenting on the possibility of
covariantizing the de Sitter Galileon beyond its fixed
 background. Similarly as in the case of the standard
 Galileon, the covariantization breaks the fundamental
  symmetry of the theory, and nothing else is expected
   here. However, one should be able to generalize this
    theory beyond any background without introducing new
     (ghost) degrees of freedom. When doing so, the first
      Galileon interaction could for instance be promoted
       to
\ba
{\mathcal{L}}^{({\rm NR})}_2 &\to& (\p \hat \pi)^2- \frac 13 R \pi\\
{\mathcal{L}}^{({\rm NR})}_3 &\to& (\p \hat \pi)^2\Box \pi-2 \pi G\mn \p^\mu \pi \p^\nu \pi-\frac 13 \mathcal{R}_{\rm GB} \pi^3
 \ea
 where $R$, $G\mn$ and $\mathcal{R}_{\rm GB}$ are respectively
 the scalar curvature, the Einstein tensor and the Gauss-Bonnet
 term in the arbitrary geometry\footnote{We point out however that
  no covariantization is unique. The most natural way to
   generalize this theory would be to start with an arbitrary
   five-dimensional metric and derive the induced Lovelock
   invariants on the brane.}. These non-minimal couplings to
   gravity could have interesting consequences of their own.
   First of all, that theory could be viable without the need
   of a separate Einstein-Hilbert.  Furthermore, these
   couplings could play a relevant role in the study of
   superluminal modes that have been know to appear around
    non-trivial spherically symmetric or cosmological backgrounds, \cite{Nicolis:2008in,Hinterbichler:2009kq}, if we took that
    couplings of the form $\pi R$ seriously beyond de Sitter.

In the very latest stage of this work we became aware that
very similar ideas have been explored by Goon {\it et. al.},
 \cite{Goon:2011qf}.

\section*{acknowledgments}

We would like to thank Andrew Tolley for useful conversations.
 This work is supported by the SNF.


\end{document}